\documentclass[twocolumn, prl, showpacs,amsmath]{revtex4}

\usepackage{graphicx,amsmath,amsfonts, amssymb,color}

\begin{document}

\title{Low-Dimensional Transport and Large Thermoelectric Power Factors in Bulk Semiconductors by Band Engineering of Highly Directional Electronic States}

\author{Daniel I. Bilc$^{1,2,*}$, Geoffroy Hautier$^3$, David Waroquiers$^3$, Gian-Marco Rignanese$^3$ $\&$ Philippe Ghosez$^{1,*}$}
\affiliation{
$^1$ Theoretical Materials Physics, Universit\'e de Li\`ege, 4000 Sart Tilman, Belgium \\
$^2$ Molecular and Biomolecular Physics Department, National Institute for Research and Development of Isotopic and Molecular Technologies, RO-400293 Cluj-Napoca, Romania \\
$^3$ Institute of Condensed Matter and Nanosciences, Universit\'e catholique de Louvain, 1348 Louvain-la-neuve, Belgium } 





\begin{abstract}

Thermoelectrics are promising to address energy issues but their exploitation is still hampered by low efficiencies. So far, much improvement has been achieved by reducing the thermal conductivity but less by maximizing the power factor. The latter imposes apparently conflicting requirements on the band structure: a narrow energy distribution and a low effective mass. Quantum confinement in nanostructures or the introduction of resonant states were suggested as possible solutions to this paradox but with limited success. Here, we propose an original approach to fulfill both requirements in bulk semiconductors. It exploits the highly-directional character of some orbitals to engineer the band-structure and produce a type of low-dimensional transport similar to that targeted in nanostructures, while retaining isotropic properties. Using first-principles calculations, the theoretical concept is demonstrated in Fe$_2$YZ Heusler compounds, yielding power factors 4-5 times larger than in classical thermoelectrics at room temperature. Our findings are totally generic and rationalize the search of alternative compounds with a similar behavior. Beyond thermoelectricity, these might be relevant also in the context of electronic, superconducting or photovoltaic applications. 
\end{abstract}

\maketitle


Thermoelectricity, realizing the direct conversion between thermal and electrical energies, is a very promising avenue for renewable energy generation. The efficiency of a thermoelectric  (TE) material can be described by its figure of merit $ZT$, defined as $ZT$=($S^2\sigma T$)/($\kappa_{e}+\kappa_{l})$, where $S$ is the thermopower, $\sigma$ the electrical conductivity, $T$ the absolute temperature, and $\kappa_{e}$ and $\kappa_{l}$ are the electronic and lattice contributions
to the thermal conductivity. 
In practice, $ZT$ should be greater than 3 for TE devices to become fully competitive with other energy conversion systems~\cite{Tritt2006, Vineis2010}.
Unfortunately, more than fifty years after the promising discovery of Bi$_2$Te$_3$-based alloys with $ZT$$\sim$1~\cite{Goldsmid1954}, increasing $ZT$ further remains a real challenge. 
Huge efforts have been dedicated to the lowering of $\kappa_{l}$ using specific crystal structures (e.g., phonon glass-electron crystals~\cite{Slack1995}) and nanostructuring~\cite{Hicks1993, Chen2003, Snyder2008, Kanatzidis2010, Kanatzidis2012}, leading to the generation of materials with $ZT$=1 -- 2.4 within the last decade~\cite{Venkatasubramanian,Harman,Hsu}. Record low values $\kappa_{l}$=0.22 -- 0.5 W/mK~\cite{Venkatasubramanian, Kanatzidis2010} were achieved and it is unlikely that these values 
can still be significantly decreased. At this stage, as emphasized by Kanatzidis~\cite{Kanatzidis2010}, the next
step forward should come from new breakthrough ideas on how to significantly enhance $S^2\sigma$, the power factor (PF). 
 
A promising avenue was proposed by Hicks and Dresselhaus~\cite{Hicks1993,Heremans2013} who predicted theoretically that quantum confinement of electrons in multiple wells can substantially increase the PF.
It was confirmed experimentally that the PF in the confined region of nanostructures~\cite{Hicks1996,Ohta2007} can indeed be larger than in related bulks. However, the gain in the confined region is partly counterbalanced by the contribution from the barrier  material producing the confinement. More recently, Mahan and Sofo~\cite{Mahan} searched for what should be the ideal
shape of the so-called ``transport distribution function'' that optimizes $ZT$.
They reached the conclusion that the best materials would combine (i) a distribution of carrier energy as narrow as possible and (ii) high carrier velocities in the direction of the applied field.
Satisfying both of these two criteria seems difficult in practice: narrow energy distributions are typically associated with flat energy bands while high carrier velocities are necessarily associated with highly dispersive bands. This could partly be achieved in rare-earth compounds like YbAl$_3$, yielding very large PF at low temperature~\cite{YbAl3}. However, those are metals with TE properties rapidly decreasing with increasing T.  Alternatively, attempts to combine these apparently incompatible requirements in semiconductors have relied on the ``band structure engineering'' of narrow energy features in the density of states from in-gap and resonant states near the band edges~\cite{Bilc2004, Ahmad2006a, Bilc2011, Heremans, Banerji}. 

Here, we show theoretically that the seemingly conflicting requirements formulated by Mahan and Sofo~\cite{Mahan} can actually be combined within the same band of certain semiconductors, exploiting the highly-directional character of some orbitals. This is achieved without any nanostructuring or introduction of resonant states.  It yields, in the bulk phase, a type of low-dimensional transport similar to that targeted by Hicks and Dresselhaus in nanostructures~\cite{Hicks1993}, while simultaneously retaining isotropic transport properties at the macroscopic level, concretizing ideas recently proposed by Parker {\it et al.}~\cite{Parker}. The concrete consequences of this finding on the TE 
  \begin{widetext}
 \begin{figure}[t]
  \begin{minipage}[t]{1.0\textwidth}
  \centering\includegraphics[width=0.9\textwidth, angle=0]{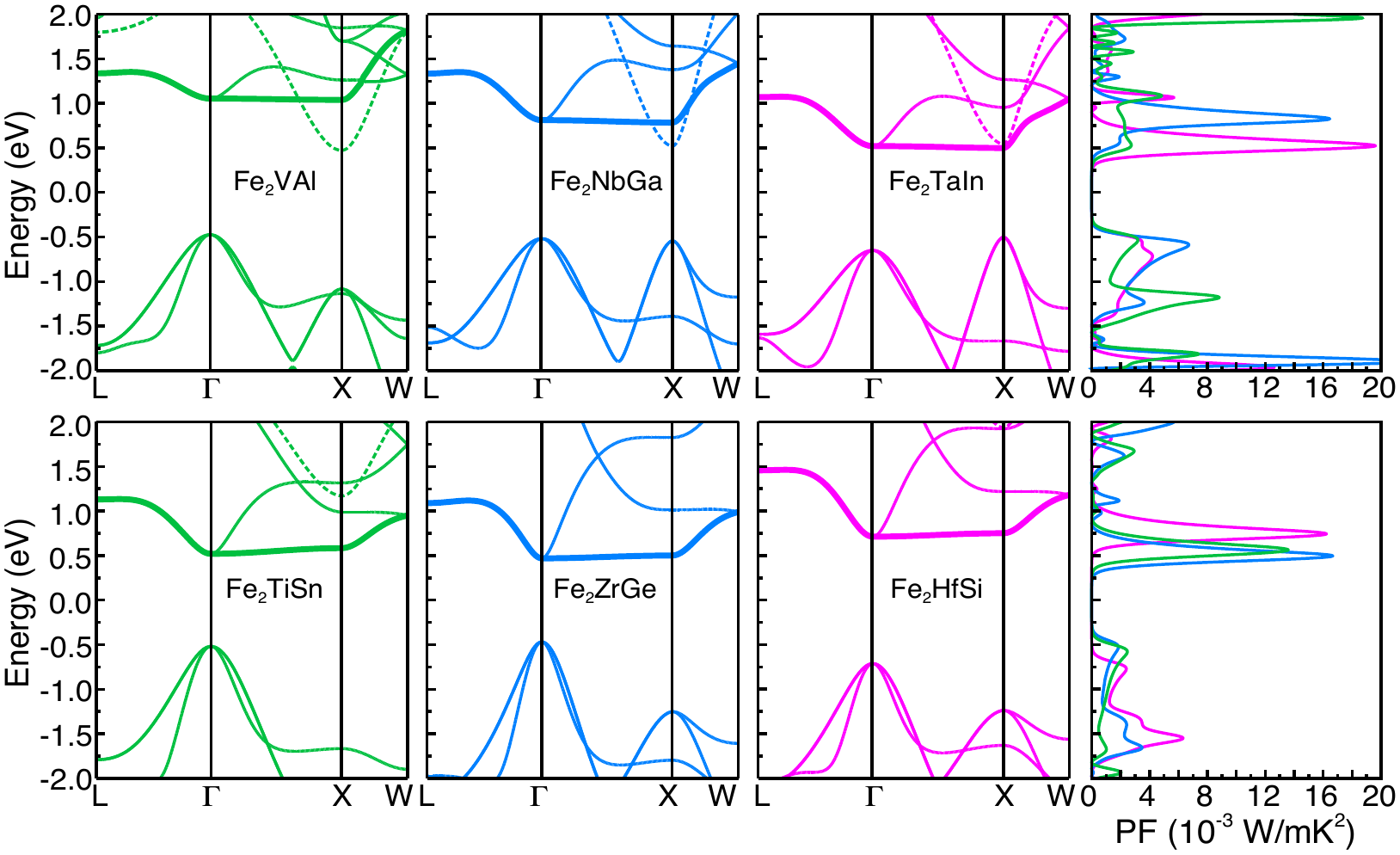}
   \caption{\label{S1} { \bf Electronic band structures and power factors (PF=$S^2\sigma$) at 300 K for the sets S1 and S2 of compounds.} 
   Fe$_2$YZ full Heusler compounds with Y=V (green), Nb (blue), Ta (magenta) and Z=Al, Ga, In in 1$^{st}$ row. Fe$_2$YZ full Heusler compounds with Y=Ti (green), Zr (blue), Hf (magenta), and Z=Sn, Ge, Si in 2$^{nd}$ row. Dotted lines indicate the highly-dispersive lowest conduction band of dominant Y~$e_g$ character. The bold lines shows the band of Fe~$e_g$ character that is flat along the $\Gamma$X direction and highly-dispersive along others.}
  \end{minipage}
\end{figure}
\end{widetext}
properties is demonstrated in the class of Fe$_2$YZ full Heusler compounds: by engineering the appearance of Fe $e_g$ states at the bottom of the conduction bands, power factors 4-5 times larger than in classical thermoelectrics (such as PbTe or Bi$_2$Te$_3$) can be obtained at room temperature. Beyond revealing the unexpected interest of a class of compounds often considered as modest thermoelectrics ~\cite{Nishino}, our results rationalize the search of alternative compounds with optimal power factor. The underlying concepts are totally generic and can be combined with other known strategies for decreasing the thermal conductivity in order to increase the TE figure of merit.

The electronic properties were studied within density functional theory using CRYSTAL~\cite{CRYSTAL}. We adopted a B1-WC hybrid functional scheme~\cite{Bilc2008}, which was previously shown accurate for describing the electronic and TE properties of this class of compounds~\cite{Bilc2011}. The electronic transport properties were studied within the Boltzmann transport formalism and constant relaxation time approximation as implemented in BoltzTraP~\cite{BoltzTraP}.  Within this approximation $S$ is independent of the relaxation time $\tau$, whereas $\sigma$ and PF depend linearly on $\tau$. The relaxation time was set to $\tau$=3.4$\times$10$^{-14}$~s in all the calculations. This value was determined by fitting the electrical resistivity $\rho$ to the experimental value of 0.65~m$\Omega$cm for Fe$_2$VAl$_{1-x}$M$_x$ (M = Si, Ge) systems at doping $x$=0.03 and 300K~\cite{Nishino, Vasundhara}.  For Fe$_2$VAl, the calculated PF is 3~mW/mK$^2$ (Fig. \ref{S1})  in close agreement with experimental data. In order to estimate with more accuracy $ZT$ at high $T$, we went beyond constant relaxation time approximation and considered $T$ and energy E dependences of the relaxation time $\tau (T,E)$ for the acoustic and polar optical phonon scattering mechanisms  (Ref.~\cite{SM}, text). The thermodynamical stability was assessed using the generalized gradient approximation (GGA) from Perdew Burke and Ernzheroff (PBE) within a plane augmented wave (PAW) approach and using VASP~\cite{VASP, PBE}. The computational parameters and pseudopotentials are similar to the ones used in the Materials Project~\cite{MP, Jain2011}.  For each chemical system (e.g., Fe-Ti-Sn), we computed the chemistry in the Heusler crystal structure but also other ternary crystal structures obtained from Heusler-forming systems. The stability of each Heusler phase was evaluated versus all phases present in the Materials Project and our generated ternary phases using the convex hull construction as implemented in the pymatgen package~\cite{Ong2013}.
  
We start our search from Fe$_2$VAl that, in spite of relatively modest TE properties ($ZT$$\sim$0.13-0.2~\cite{Nishino, Mikami2012}, PF = 4-6~mW/mK$^2$~\cite{Nishino, Vasundhara} at 300-400 K), is considered for low-cost TE applications~\cite{Mikami2009}. 
As clarified recently~\cite{Bilc2011}, it is an intrinsic semiconductor, with a low band gap between the highest valence bands of dominant Fe $t_{2g}$ character and a highly-dispersive lowest conduction band of dominant V~$e_g$ character (see dotted line in Fig. \ref{S1}). Interestingly Fe$_2$VAl also exhibits a ``flat-and-dispersive'' band of Fe~$e_g$ character that is very flat along the $\Gamma$X direction of the Brillouin zone (BZ) and highly dispersive along others (see bold line in Fig. \ref{S1}). This band combines the above-mentioned features identified by Mahan and Sofo to produce a large PF. It lies however $\sim$0.6 eV above the bottom of the conduction band, and is not active in transport at room temperature for optimal doping at electron concentrations $n$$\sim$10$^{19}$~cm$^{-3}$. In order to move its position towards the bottom of the conduction band we performed atomic substitutions at Y and Z sites.

We consider in Fig. \ref{S1} a first set (S1) of Fe$_2$YZ full Heusler compounds with Y=V, Nb, Ta and Z=Al, Ga, In. Going from 3$d$ to 5$d$ transition metal elements at the Y site tends to move the Y~$e_g$ dispersive band upwards. Furthermore, going to higher-mass elements at Y and Z sites increases the lattice parameter (and the Fe-Fe distance), resulting in a decrease of Fe-Fe interactions lowering the Fe~$e_g$ levels (Ref.~\cite{SM}, text). Consequently, in many compounds, the Fe~$e_g$ flat-and-dispersive band appears close to the bottom of the conduction band. In line with our expectations, this increases the PFs at 300K from $\sim$3~mW/mK$^2$ for
Fe$_2$VAl to $\sim$12~mW/mK$^2$ for Fe$_2$NbAl and Fe$_2$TaAl and up to 16-20~mW/mK$^2$ for Fe$_2$NbGa, Fe$_2$TaGa, Fe$_2$NbIn and Fe$_2$TaIn (Fig. \ref{S1}, and Ref.~\cite{SM}, Fig. 4). Band-by-band analysis for  Fe$_2$TaIn  shows that most of the PF (90 $\%$) comes from the Fe~$e_g$ flat-and-dispersive band (Ref.~\cite{SM}, Fig. 6). 
     
In a second set (S2) of Fe$_2$YZ compounds with Y=Ti, Zr, Hf, and Z=Si, Ge, Sn, the dispersive Y~$e_g$ band is pushed up even higher in energy. It appears well above the Fe~$e_g$ bands, that are now the only lowest conduction bands (Fig. \ref{S1}, and Ref.~\cite{SM}, Fig. 5). These compounds exhibit also extremely large PFs of~$\sim$14-17~mW/mK$^2$ at 300K, consistently with what was reported in Ref.~\cite{Yabuuchi}. This time, the large PF is almost entirely generated by the Fe~$e_g$ flat-and-dispersive band ($\sim$93~$\%$ of the total PF for Fe$_2$TiSn (Ref.~\cite{SM}, Fig. 6).

In all S1 and S2 compounds exhibiting enhanced TE properties, the PF remains substantial (i.e. keeps 90\% of its peak value) in a similar and relatively wide range of carrier concentrations ($n \approx 4 \times$10$^{20}$~-~3$\times$10$^{21}$~cm$^{-3}$) at which $|S|$$\sim$150-200~$\mu$V/K. These compounds then exhibit large $S$, comparable to those of Fe$_2$VAl and the best classical thermoelectrics, but at $n$ and $\sigma$ values that are about one order of magnitude larger (Ref.~\cite{SM}, Fig. 7).

\begin{widetext}
 \begin{figure}[t]
  \begin{minipage}[t]{1.0\textwidth}
  \centering\includegraphics[width=0.95\textwidth, angle=0]{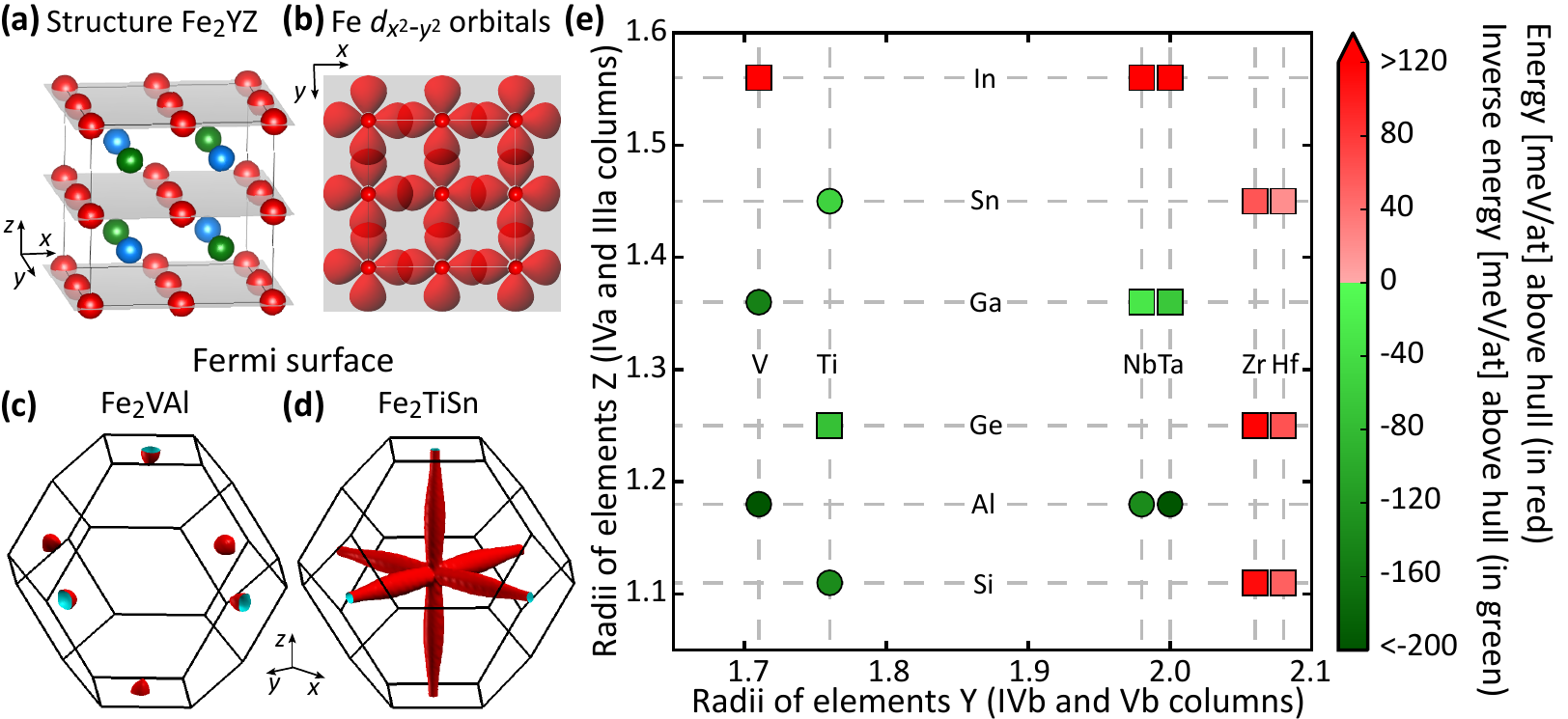}
   \caption{\label{Structure} {\bf Atomic structure, Fermi surfaces, and thermodynamical stability of
Fe$_2YZ$ compounds. (a)} Full Heusler structure consisting of four
interpenetrating fcc lattices with Fe, $Y$ and $Z$ atoms represented in red,
green and blue. It can also be viewed as alternating Fe$_2$ and $YZ$
planes along the \{001\} (see Fe$_2$ planes highlighted in grey),
\{100\}, and \{010\} directions. {\bf (b)} Sketch of the
Fe~$d_{x^2-y^2}$~$e_g$ orbitals overlapping in \{001\}-Fe$_2$ planes.
{\bf (c,d)} Fermi surfaces of Fe$_2$VAl and Fe$_2$TiSn at doping
concentration yielding the maximum PFs at 300K.
{\bf (e)} Map of the thermodynamical stability (as measured by the energy
with respect to hull) computed at 0 K as a function of the $Y$ and $Z$ atomic
radii. A negative number (inverse energy above hull, green color)
corresponds to a stable phase and a positive number (energy above hull,
red color) to an unstable one. Circles (resp. squares) correspond to
compounds (resp. not) previously synthesized. }
  \end{minipage}
\end{figure}
\end{widetext}

Taking Fe$_2$TiSn as a representative example, the inspection of the carriers contributing to the transport properties at 300K  provides further insight into the enhancement of the PFs respect to Fe$_2$VAl (Fig. \ref{Structure}c and d). In the latter, the sizable TE properties at optimal doping arise from electrons located in small pockets centered at $X$ and associated to highly-dispersive V~$e_g$ bands (Fig. \ref{Structure}c). In Fe$_2$TiSn, the enhanced transport properties at optimal doping are produced instead by electrons from the Fe~$e_g$ lowest-conduction states located in three orthogonal tubes extending along $\Gamma-X$ directions and intersecting at $\Gamma$ (Fig. \ref{Structure}d). These tubes can be viewed as the juxtaposition along the entire $\Gamma-X$ direction of consecutive electron pockets similar to those of Fe$_2$VAl, so explaining the improved transport properties. This is an optimal realization of the concept of pocket engineering brought forward by Snyder~\textit{et al.}~\cite{Pei} where large $S$ and $\sigma$ values are obtained without compromising mobility by introducing degenerate low effective mass pockets in the Brillouin Zone.

In each cartesian direction, the Heusler Fe$_2$YZ structure can be seen as made of Fe$_2$ (alternating with YZ) atomic planes. For clarity, only the family of \{001\}-Fe$_2$ planes is illustrated in Fig. \ref{Structure}a but similar families can be drawn along the two other cartesian directions. The tubular shape of the Fermi surface of Fe$_2$TiSn and related compounds is originating from the highly-directional character of the Fe~$e_g$ orbitals. The conduction states of the three tubes in Fig. \ref{Structure}d are associated to the flat band along $\Gamma$X (and symmetrically equivalent $\Gamma$Y and $\Gamma$Z directions) in Fig. 1.  For the tube along $z$, these states are made of Fe~$d_{x^2-y^2}$ orbitals (see Ref.~\cite{SM} Fig. 3)  which strongly overlap in \{001\}-Fe$_2$ planes (strong $\sigma^*$ bonds along $x$ and $y$, Fig. \ref{Structure}b) but do not interact significantly from plane to plane  (weak $\delta^*$ bonds along $z$). The same is true, {\it mutatis mutandis}, for the tubes along $x$ and $y$. This anisotropy of the orbital interactions gives rise to electronic bands that are highly-dispersive in two directions (m$_{t}$$\sim$0.3 m$_e$ for Fe$_2$TiSn and m$_{t}$$\sim$0.2 m$_e$ for Fe$_2$TiSi) and flat in the third one (m$_{l}$$\sim$26 m$_e$ for Fe$_2$TiSn and m$_{l}$$\sim$90 m$_e$ for Fe$_2$TiSi), so concretizing the ideas of Mahan and Sofo~\cite{Mahan}. The tubular shape of the Fermi surface comes from the large effective mass ratio (R=m$_{l}$/m$_{t}$$\sim$87 for Fe$_2$TiSn and R$\sim$450 for Fe$_2$TiSi) and highlights that these compounds exhibit a kind of two-dimensional electronic transport in Fe$_2$ planes~\cite{Note}, similar to what was proposed in nanostructures in order to realise increased TE performance~\cite{Hicks1993}. Here however, it is achieved in bulk cubic compounds. This yields a periodic repetition of two-dimensional conductive channels (i.e. Fe$_2$ planes)  at the ultimate unit-cell scale. Moreover, the isotropic character of the properties is preserved through the coexistence of symmetry equivalent families of \{001\}-, \{010\}- and \{100\}-Fe$_2$ planes. In practice however, all the conductive electrons  are not similarly contributing to
 transport in a given direction: for instance, only the states in the tubes along $x$ and $y$ (\{100\}- and \{010\}-Fe$_2$ planes) contribute significantly to the transport along $z$. All this is a concrete illustration that low-dimensional electronic structures can occur in high-symmetry cubic systems, as also proposed recently by Parker~\textit{et al.}~\cite{Parker} in another class of compounds.

From the practical point of view, a central issue concerns the thermodynamical stability of the S1 and S2 compounds. This was investigated at the first-principles level (Ref.~\cite{SM}, text).  The results are summarized in Fig. \ref{Structure}e. Stable and unstable phases are associated respectively with green and red colors. Compounds previously synthesized are stable. Fe$_2$TiGe and Fe$_2$NbGa with good TE performance are predicted to be fully stable. We see that including elements like In, Zr or Hf with large ionic 
 radii tend to destabilize the Heusler structure, although the synthesis of Fe$_2$HfSn might stay experimentally accessible. Another practical issue concerns the appearance of anti-site defects  in these Heusler compounds, which are detrimental to the TE performance~\cite{Bilc2011}. From our calculations in Ref.~\cite{SM} Table 2, S2 compounds appear to be less prone to form anti-site defects than Fe$_2$VAl.

The existence of several isostructural stable compounds with very attractive PF is an advantage for TE applications. Combining such compounds in solid solutions should allow the reduction of $\kappa_{l}$ \cite{Kanatzidis2012} while preserving the shape of their electronic band structure at the conduction band bottom (nearly identical since dominated by the same Fe $e_g$ states, see Ref.~\cite{SM} Figure 5) and the related large PF. Such a reduction of $\kappa_{l}$ has been demonstrated in Fe$_2$V$_{1-x}$W$_{x}$Al alloys~\cite{Mikami2012}, reaching values of~$\sim$3 W/mK.  The presence of heavier elements than V and Al is likely to lower $\kappa_{l}$ of the new Heusler candidates compared to Fe$_2$VAl. In this context, reasonably low values can be expected, for instance, in Fe$_2$TiSn and its solid solutions (e.g., Fe$_2$TiSn$_{1-x}$Si$_{x}$).
With $\kappa_{l}$=3.5 W/mK, we estimate $ZT$ values larger than 1 for Fe$_2$TiSn  in the temperature range between 600-900K (Ref.~\cite{SM}, Fig. 10). Such large $ZT$ values are achieved at carrier concentrations only slightly lower than those corresponding to the optimum PF and at which $S$ and $n$ remain large. These $ZT$ values are significantly larger than those predicted for Fe$_2$VAl, which properly reproduce the experiment ($ZT$ = 0.2 at 400K with $\kappa_{l}$=3.3 W/mK, in agreement with Ref. ~\cite{Mikami2012}). This attests that the beneficial effect on the PF, achieved through the low-dimensional transport in Fe$_2$TiSn, can be accompanied by a significant increase of $ZT$, without being systematically counter-balanced by another detrimental effect. 

The excellent TE properties of some of the Fe$_2$YZ compounds demonstrated theoretically in this work, combined with the low-cost and wide availability of their constitutive elements, make them very attractive for large-scale TE applications, well beyond what could have been anticipated from previous studies of Fe$_2$VAl. Going further, our results also highlight that, contrary to the current belief, extremely large PF can be intrinsic to bulk semiconductors. Our work rationalizes how this can be achieved in practice through the engineering of highly-directional states at the bottom of the conduction bands, yielding low-dimensional transport. This calls for the search of alternative families of compounds realizing the same ideas. The link that we establish with some basic requirements on the electronic band structure is particularly relevant to succeed identifying such compounds in the current context of emergent high-throughput search of alternative thermoelectrics ~\cite{HT1,HT2} that requires clear and simple design rules. Our findings might also attract interest well beyond the field of thermoelectrics: the singular Fermi surface in these systems present close similarities with those of Fe-based superconductors~\cite{Supra} and the associated low-dimensional transport might be relevant in the context of some electronic \cite{Chen} or photovoltaic \cite{Pebano} applications.

We thank P. Jacques, J.-P. Issi and N. Bristowe for useful  discussions. Work supported by EnergyWall project CoGeTher, ARC project TheMoTherm, FNRS project HiT4FiT, and a collaboration between WBI and the Romanian Academy of Sciences. Ph.G. thanks a Research Professorship of the Francqui Foundation,  D.I.B. the Romanian National Authority for Scientific Research, CNCS-UEFISCDI (Grant No. PN-II-RU-TE-2011-3-0085),  G. H. and G.-M. R. the F.R.S.-FNRS  and G. H. the European Marie Curie CIG (Grant No. HTforTCOs PCIG11-GA-2012-321988). Calculations were performed at C\'eci HPC Center, funded by F.R.S.-FNRS.$^*$Correspondence should be addressed to D.I.B. (Daniel.Bilc@itim-cj.ro) and Ph.G. (Philippe.Ghosez@ulg.ac.be).%


\begin{thebibliography}{99}
\bibitem{Note} The 1D-shape of Fermi surface is associated to 2D transport in real space. Due to the orientation of $d_{x^2-y^2}$ orbitals, the latter can be viewed as the combination of two orthogonal transport channels along the Fe-Fe bonds.
\end{thebibliography}
\end{document}